\documentstyle[12pt]{article}
\begin{document}
\baselineskip 1.5pc
\def\ii{\'{\i}}
\def\bb{$\beta\beta_{2\nu}$}
\def\nd{$^{150}Nd$}
\def\sm{$^{150}Sm$}
\bigskip
\bigskip
\centerline{\bf\large Double beta decay to excited states in \nd }
\vskip 1cm
\centerline{Jorge G. Hirsch}
\centerline{\tenit Departamento de F\'{\i}sica, Centro de
Investigaci\'on y de Estudios Avanzados del IPN,}
\centerline{\tenit A. P. 14-740 M\'exico
07000 D.F.}

\centerline{Octavio Casta\~nos and Peter O. Hess}
\centerline{\tenit Instituto de Ciencias Nucleares, Universidad Nacional
Aut\'onoma de M\'exico,}
\centerline{\tenit A. P. 70-543 M\'exico 04510 D.F.}

\centerline{Osvaldo Civitarese}
\centerline{\tenit Departamento de F\ii sica, Universidad Nacional de La
Plata,}
\centerline{\tenit c.c.67 1900, La Plata, Argentina}
\bigskip
\bigskip
\bigskip
\centerline{\bf{Abstract:}}
 The pseudo SU(3) model is used to study the double
 beta decay of
\nd ~to the ground and
excited states of \sm . Low lying collective excitations of \sm ~and its
BE(2) intensities are well reproduced. Expressions for the two neutrino
double beta decay to excited
states are developed
and used to describe the decay of \nd . The existence of selection rules
which strongly
restrict the decay is discussed.
\vfill
\eject
\bigskip
\bigskip
\centerline{\bf{Introduction}}
\bigskip
\bigskip

The neutrinoless double beta decay ($\beta\beta_{0\nu}$),
undetected up to now,
provides the more stringent limits to the Majorana
mass of the neutrino $<~m_{\nu_ e}~> \le 1.1 eV$\cite{Mai94}.
Its detection would imply an indisputable evidence of physics beyond the
standard model and would be useful in order to select Grand
Unification Theories\cite{Ver86}.

Theoretical nuclear matrix elements are needed to convert
experimental half-life limits, which are available for many
$\beta\beta$-unstable isotopes\cite{Moe93}, into constrains for
particle physics parameters such as the effective Majorana mass of the
neutrino and the contribution of right-handed currents to the weak
interactions. Thus, these matrix elements are essential to understand the
underlying physics.

The two neutrino mode of the double beta decay (\bb ) is
allowed as a second order process in the standard model. It has
been detected in nine nuclei\cite{Moe93} and has served as a
test of a variety of nuclear models.
The calculation of the \bb ~and $\beta\beta_{0\nu}$ matrix elements
requires different theoretical methods. Therefore a successful prediction
of the former cannot be considered a rigorous test of the latter, but gives
some confidence.
However it is the best available proof we can impose to a nuclear model
used to predict the $\beta\beta_{0\nu}$ matrix elements.

Many experimental groups have reported measurements of \bb ~
processes\cite{Moe93}. Nearly for all the cases the ground state (g.s.) to
ground
state ($0^+ \rightarrow 0^+$) decay was investigated. In direct-counting
experiments the analysis of the sum-energy spectrum of the emitted
electrons allowed the identification of the different $\beta\beta$-decay
modes\cite{Pie94}.

Recently the possibility of detecting $\beta\beta$ decay into excited
states of the daughter nucleus by measuring the gamma radiation has
exerted some attraction among experimentalists. This
is due to the fact that
phase space integrals scale as the energy available for the decay and
consequently decrease for excited states. In the case of the decay to a first
excited $2^+$ state
the phase space factor contains terms which are
antisymmetric in the energies of
the two outgoing electrons and antineutrinos, resulting in a large reduction
of the corresponding integral. It makes such transitions very
difficult to observe\cite{Doi85,Moe94}. Also they are inclusive experiments
which cannot distinguish between the
different $\beta\beta$ decay modes. On the other hand the
detection of a gamma ray  gives a much clearer signal
than a continuous electron spectrum as in the case of the
$g.s. \rightarrow g.s.$ decay.

The pioneer work of Bellotti {\em et. al}\cite{Bel82} determined
lower limits to
the $\beta\beta$ decays to excited states of six nuclei.
Lower limits for the decay of
$^{128}Te$ and $^{130}Te$ to the first $2^+$ state of the corresponding
Xenon isotopes have been reported\cite{Bel87}.
The $\beta\beta$ decay of $^{76}Ge$ to excited states of $^{76}Se$ was
studied looking for the detection of one or two photons in coincidence
with two electrons\cite{Mor91,Bec92}.
The half-lives for the \bb ~decay from the g.s. of $^{100}Mo$, $^{96}Zr$
and \nd ~to the first
excited $0^+$ state of the daughter nuclei have been estimated for the first
time in \cite{Bar90}
assuming that the nuclear matrix elements of the transitions to the g.s.
and excited $0^+$ state are the same.
Experimental studies of the decay to excited states looking for the
$\gamma$ ray signature have been performed for $^{96}Zr$\cite{Arp94},
$^{116}Cd$\cite{Pie94,Bar90a,Dan93}, and
$^{100}Mo$\cite{Bar90a,Nem92,Kud92,Bar92}.
The detection of the \bb ~to the first excited $0^+$ state
was reported for the first time in \cite{Bar92,Bar91}
The factibility of studying the \bb ~to excited states in \nd ~has been
discussed in recent years\cite{Pie94,Bar90,Arp94,Nem94}. In
\cite{Arp94} preliminary results were reported.

Theoretical analysis of the \bb ~to excited states have been performed
in the
context of the QRPA formalism for $^{100}Mo$\cite{Gri92,Suh94},
$^{136}Xe$\cite{Suh93,Civ94} and $^{76}Ge$\cite{Civ94a,Civ94b},
$^{116}Cd$\cite{Pie94} and also for
$^{82}Se,^{110}Pd$ and $^{128,130}Te$\cite{Sto94}.
When the first excited $0^+$ state was studied
it was assumed that it
is a member of a two quadrupole-phonon triplet.
The QRPA calculation for $^{100}Mo$ exhibits an overestimation of the
amplitude of the
$\beta\beta_{2\nu}$ decay to this excited state when the decay to the
ground state is reproduced\cite{Gri92,Suh94}. For the case of
$^{116}Cd$\cite{Pie94} the matrix element of this decay is five
times greater than that associated with the decay to the ground state.

Although there is no reported calculation of the \bb ~of \nd ~to excited
states of \sm , this nucleus was mentioned as a suitable
candidate for this decay. In \cite{Bar90} this
conclusion has been reached assuming that the nuclear matrix
elements for the \bb ~to
the ground state and the first excited $0^+$ state are equal. In
\cite{Pie94} it was speculated that if the matrix element for the \bb ~of
\nd ~would show a similar enhancement over that of the g.s. decay as found
for $^{116}Cd$, the decay rate into this excited level could even exceed
that of the g.s. decay.

In the present paper we perform an analysis of the \bb ~decay of \nd
{}~to excited $0^+$ and $2^+$ states of \sm ~using the pseudo SU(3) formalism.
This theoretical model is well suited to describe the collective spectra.
Under this scheme the first $2^+$ and $4^+$ states are
members of the g.s. rotational band, while
the second excited $0^+$ and $2^+$ states are members of an excited rotational
band.
The third excited $0^+$ state is the head of another rotational band.
We will show that whithin the pseudo SU(3) model
the \bb ~decay to the first excited $0^+$ state will be
cancelled  while the other decays will be strongly suppressed.
The necessary formalism to study the \bb ~decay to $2^+$ states is also
developed.

In Section 2 the pseudo SU(3) formalism and the model Hamiltonian are
briefly reviewed. In Section 3 the summation method is used to obtain the
\bb ~matrix elements for the decay to the $2^+$ states. Section 4
contains the explicit formulae needed to evaluate the \bb ~matrix elements
in the pseudo SU(3) scheme.
The nuclear structure analysis of \sm ~ is given in
Section 5. In Section 6 the
\bb ~nuclear matrix elements and half-lives are presented. Conclusions are
drawn in the last Section.

\bigskip
\bigskip
\centerline{\bf The pseudo SU(3) formalism}
\bigskip
\bigskip

In order to obtain a microscopical description of the low lying energy states
of \nd ~and \sm ~we will use the pseudo SU(3) model
which successfully  describes collective excitations in rare earth nuclei and
actinides\cite{Dra84} as well as the $g.s. \rightarrow g.s.$ \bb ~and
$\beta\beta_{0\nu}$ decays
of six heavy deformed nuclei\cite{Cas93,Cas94,Hir94}.

In the pseudo $SU(3)$ shell model coupling scheme\cite{Rat73} normal
parity orbitals $(\eta ,l,j)$ are identified with orbitals of a harmonic
oscillator of one quanta
less $\tilde \eta = \eta-1$.  This set of orbitals with $\tilde j
= j = \tilde  l + \tilde s$,  pseudo spin $\tilde s =1/2$ and  pseudo
orbital angular momentum  $\tilde l$
define the so called pseudo space. Recently it was found an
analytic expression for the transformation of the normal
parity orbitals to the pseudo space \cite{Cas92}.  Applying this
transformation to the spherical Nilsson Hamiltonian it can be
shown explicitly that the strength of the pseudo spin orbit
interaction is almost zero for heavy nuclei and the orbitals
$j = \tilde l \pm 1/2$ are nearly degenerate. For configurations
of identical particles occupying a single j orbital of abnormal parity
a convenient characterization of states is made by means of the seniority
coupling scheme.

The many particle states of $n_\alpha$ nucleons in a given shell
$\eta_\alpha$,
$\alpha = \nu $ or $\pi$, can be defined by the totally antisymmetric
irreducible representations
$\{ 1^{n^N_\alpha}\} $ and $\{1^{n^A_\alpha}\}$ of unitary groups.
The dimensions of the normal $(N)$ parity space is
$\Omega^N_\alpha = (\tilde\eta_\alpha + 1) (\tilde\eta_\alpha +2)$ and
that of the unique $(A)$ space is $\Omega^A_\alpha =
2\eta_\alpha +4$, with the constraint
$n_\alpha = n^A_\alpha	+ n^N_\alpha$.
Proton and neutron states are coupled to angular momentum $J^N$ and $J^A$
in both the normal and unique parity sectors, respectively. The
wave function of the many-particle state  with angular
momentum $J$ and projection $M$ is expressed as a direct product of the
normal	and unique parity ones, as:

\begin{equation}
|J M > = \sum\limits_{J^N J^A} [|J^N> \otimes |J^A>]^J_M
\end{equation}
We are interested in the low energy states of \nd ~ and \sm ~which
have $J=0,2,3,4,6$.

For even-even heavy nuclei it has been shown that if the
residual neutron-proton interaction is of the quadrupole type,
regardless of the interaction in the proton and neutron
spaces, the most important
normal parity configurations are
those with highest spatial symmetry $\{ \tilde f_\alpha \} =
\{ 2^{n^N_\alpha /2}\}$\cite{Dra84}.
This statement is valid for yrast states below the backbending region.
This  implies that $ \tilde S_\pi = \tilde S_\nu = 0$,
i.e. only pseudo spin zero configurations are considered.

Additionally in the abnormal parity space only seniority zero
configurations
 are taken into account.  This simplification implies that
 $J^A_\pi = J^A_\nu = 0$. This is a very
strong assumption quite useful in order to simplify the calculations.
Its effects upon the present calculation are discussed below.

The double beta decay, when described in the pseudo SU(3) scheme, is
strongly dependent on the occupation numbers for protons and neutrons in
the normal and abnormal parity states $n^N_\pi, n^N_\nu, n^A_\pi,
n^A_\nu$\cite{Cas93,Cas94}.
These numbers are determined filling the Nilsson levels from below, as
discussed in \cite{Cas93,Cas94}.
In particular the \bb ~decay is allowed only if
it fulfils the following relationships

\begin{equation}
\begin{array}{l}
n^A_{\pi ,f} = n^A_{\pi ,i} + 2~~,
\hspace{1cm}n^A_{\nu ,f} = n^A_{\nu ,i}~~, \\
n^N_{\pi ,f} = n^N_{\pi ,i}~~ ,
\hspace{1.7cm} n^N_{\nu ,f} = n^N_{\nu ,i} - 2 ~~.\label{num}
\end{array}
\end{equation}

For \nd ~it was assumed a deformation $\beta \approx 0.28$\cite{Ram87}.
We have  obtained the occupation numbers

\begin{equation}
n^A_\pi = 4~~,~~~ n^N_\pi =6~~,~~~
n^A_\nu = 2 ~~,~~~ n^N_\nu = 6~~.     \label{numnd}
\end{equation}

For \sm  ~it is found that $\beta \approx 0.19$\cite{Ram87} but we were
forced to select a higher deformation to satisfy relations ( \ref{num}).
According to \cite{Sak84}
this higher deformation is more appropriate for $^{152}Sm$ and is related
with some departure from a rotational behavior in the ground state band
of \sm . The selected occupation numbers for \sm ~ are
\begin{equation}
n^A_\pi = 6~~,~~~ n^N_\pi =6~~,~~~
n^A_\nu = 2 ~~,~~~ n^N_\nu = 4~~.  \label{numsm}
\end{equation}

In order to analyze the spectra and transitions amplitudes of \sm ~ we have
selected the standard version of the pseudo SU(3) Hamiltonian\cite{Dra84}.
It is constructed by a spherical central potential, a quadrupole-quadrupole
interaction and a residual force.  The latter allows the fine tuning of
low lying spectral features like $K$ band splitting and the effective
moments of inertia. The hamiltonian looks

\begin{equation}
H = \sum_{\alpha}{H_\alpha - {1 \over 2} \chi \bf{Q}^a\cdot \bf{Q}^a}
\zeta_1 \; K^2 + \zeta_2 \; L^2 . \label{hamil}
\end{equation}

The spherical Nilsson hamiltonian which describe the single-particle motion
of neutrons ($\alpha = \nu$) or protons ($\alpha = \pi$) is:

\begin{equation}
 H_\alpha = \sum_{s}{\hbar \omega \left\{ \eta_{\alpha s}
	+ {3 \over 2}  -2 k_\alpha \
	   {\bf\vec{L}_{\alpha s} \cdot \vec{S}_{\alpha s}} - k_\alpha
		\mu_\alpha {L}^{2}_{\alpha s} \right\} - {\rm V_\alpha}}
=  \sum_{s,\alpha} \epsilon_{s\alpha} a^{\dagger}_{s\alpha} a_{s\alpha}
\label{nilsson} \end{equation}

\noindent
where $\eta  = \tilde\eta + 1$ denotes the harmonic oscillator
number operator and $\hbar \omega$ determines the size of the shell.
A constant term ${\rm V_\nu} \ ({\rm V_\pi})$ is included
which represents the depth of the neutron (proton) potential well.
In (6) the second quantization representation of $H_\alpha$ is given,
$\epsilon_{s\alpha}$ being the single-particle energies.

The quadrupole operator
${\bf {Q}^a}= \sum_{s}{\{{q_\pi}_s  + {q_\nu}_s\}}$
acts only within a shell and do not mix different shells. The residual
interaction, $K^2$, is a linear combination of	${L}^{2}$,
${X}_{3}$ and ${X}_{4}$, defined as
\begin{equation}
\begin{array}{ll}
 {L}^{2} =& \sum_{i}^{3}{{L}_{i}^2} \\
 {X}_{3} = & \sum_{i,j}^{3}{{L}_{i}{Q}_{ij}^a {L}_{j}} \\
 {X}_{4} = & \sum_{i,j,k}^{3}{{L}_{i} {Q}_{ij}^a {Q}_{jk}^a {L}_{k}} \\
\end{array}
\end{equation}
\noindent
They are rotational invariant and scalar operators built by
generators of the algebra of $SU(3)$\cite{Dra84,Cas94} and
${L}_{i}$ and ${Q}_{ij}^a$ are cartesian forms of the total
angular
momentum and the quadrupole operators, respectively. The $K$ is interpreted
to be the third component of the total angular momentum on an intrinsic
body-fixed symmetry axis of the system, which is given by
\begin{equation}
                {K}^{2} = ({ \lambda}_{1} { \lambda}_{2} {L}^{2} +
                 { \lambda}_{3} {X}_{3} + {X}_{4}) / ( 2
                 { \lambda}_{3}^2 + { \lambda}_{1}
                 { \lambda}_{2})
\end{equation}
with the parameters $\lambda_i$ denoting the eigenvalues of the mass
quadrupole operator, which are related to the $SU(3)$ labels $(\lambda,\mu)$
through the expressions
\begin{equation}
                \lambda_1 = {1 \over 3} (\mu-\lambda) , \ \
                \lambda_2 = -{1 \over 3} (\lambda + 2 \mu + 3), \ \
                \lambda_3 = {1 \over 3} (2 \lambda +  \mu + 3) .
\end{equation}

Although the quantum number $K$ used to define the orthonormalised basis
is not the same as the Elliot $\kappa$ the states studied in the present
work satisfy quite accurately the relationships:
\begin{equation}
K^2 | K=1 > = 0 ,\ \ K^2 | K=2 > \approx 4 | K=2 > \ .
\end{equation}

It would be possible to add the the hamiltonian (\ref{hamil}) terms which
distinguish different irreps. For the sake of simplicity we keep this
simplest version.

With the occupation numbers determined in  Eq (\ref{numnd}) and
(\ref{numsm}) and the hamiltonian (\ref{hamil}) the wave function of
the deformed ground state of \nd ~can be written\cite{Cas93,Cas94}

\begin{equation}
\begin{array}{ll}
|^{150}Nd, 0^+\rangle \equiv &|0^+_i \rangle =
 | \ (h_{11/2})^4_\pi,	\ J^A_\pi = M^A_\pi = 0; \
(i_{13/2})^2_\nu ,\ J^A_\nu = M^A_\nu = 0 >_A \\
& | \ \{ 1^6\}_\pi  \{2^3\}_\pi (12,0)_\pi;
\ \{ 1^{6}\}_\nu \{2^3\}_\nu (18,0)_\nu; \ 1 (30,0) K=1 J = M = 0 >_N
\ \ ,  \label{nd}
\end{array}
\end{equation}

\noindent
and the deformed low energy states of \sm ~ are described by
the wave functions

\begin{equation}
\begin{array}{ll}
|^{150}Sm, J^+_\sigma  \rangle \equiv &| J^+_\sigma  \rangle =
 | \ (h_{11/2})^6_\pi,	\ J^A_\pi = M^A_\pi = 0; \
(i_{13/2})^2_\nu ,\ J^A_\nu = M^A_\nu = 0 >_A \\
 & | \ \{ 1^6\}_\pi  \{2^3\}_\pi (12,0)_\pi;
\ \{ 1^{4}\}_\nu \{2^2\}_\nu (12,2)_\nu; \ 1 (\lambda ,\mu )_\sigma K~ J~
M >_N \ ,  \label{sm}
\end{array}
\end{equation}

\noindent
where $J^+_\sigma$ denotes a state with angular momentum $J$, positive
parity and associated with the SU(3) irrep $(\lambda ,\mu )_\sigma$.
In this approach we are assuming that the first $0^+, 2^+, 4^+$
states of \sm ~are the low energy sector of a rotational band described by
the normal
$(\lambda ,\mu )_{g.s.} = (24,2)$ strong coupled pseudo SU(3) irrep,
the second $0^+, 2^+$ states belong to a second rotational band with
$(\lambda ,\mu )_1 =
(20,4)$, and the third $0^+$ state is the head of another rotational band
described by the pseudo SU(3) irrep $(\lambda ,\mu )_2 = (22,0)$. We will
discuss also a gamma band associated with $(\lambda ,\mu)_{g.s.}$ with K=2.

\bigskip
\bigskip
\centerline{\bf The \bb ~decay to excited $2^+$ states}
\bigskip
\bigskip

The inverse half life of the two
neutrino mode of the $\beta\beta$-decay
can be cast in the form\cite{Doi85}

\begin{equation}
	\left[\tau^{1/2}_{2\nu}(0^+ \rightarrow J^+_\sigma)\right]^{-1} =
      G_{2\nu}(J^+_\sigma) \ | \ M_{2\nu}(J^+_\sigma) \ |^2 \ \ .
\end{equation}

\noindent
where $G_{2\nu}(J^+_\sigma)$ are kinematical factors.
They depend on $E_{J;\sigma} = {1 \over 2} [{\it Q}_{\beta  \beta}-
E(J_\sigma )] + m_e c^2$ which is the half of total energy released. The
nuclear matrix element is

\begin{equation}
M_{2\nu}(J^+_\sigma) \approx M_{2\nu}^{GT}(J^+_\sigma) =
{1 \over{\sqrt{J+1}}}
\sum_{N}{}
{{ \langle  J^+_\sigma \ || \, \Gamma \, || \ 1^+_N \rangle \  \langle1^+_N \
	|| \,\Gamma \, ||\  0^+_i \rangle \,}
	\over{\mu_N}^{J+1}}  \label{m1}
\end{equation}
with the Gamow-Teller operator $ \Gamma$ expressed as
\begin{equation}
 \Gamma_m = \sum_s \ \sigma_{ms} t^-_s \equiv \sum_{\pi \, \nu}
 \sigma (\pi ,\nu ) [a^{\dagger}_{\eta_{\pi} l_{\pi} \, {1 \over 2};j_{\pi}}
\otimes \tilde a_{\eta_{\nu} l_{\nu} \, {1 \over 2};j_{\nu}}]^1_m
\hspace{1cm}m=1,0,-1.	\label{gt}
\end{equation}
  The  energy denominator is $\mu_N = E_{J,\sigma } + E_N -E_i$  and it
contains the intermediate $E_N$  and initial $E_i$ energies. The kets
$|1^+_N\rangle$  denote intermediate states.

The mathematical expressions needed to evaluate the nuclear matrix
elements of the $g.s. \rightarrow g.s.$ \bb ~decay
in the pseudo SU(3) model were developed
recently\cite{Cas93,Cas94}. The same formulae describe the
 decay to the first excited $0^+$ state by replacing the
values of the strong coupled
irrep  $(\lambda ,\mu )_{g.s.}$ of Eq. (\ref{sm}) with those
corresponding to excited bands.

We will concentrate first on the derivation
of the matrix element $M_{2\nu}(2^+_\sigma)$  to the $2^+$ excited states.
The formulae for this decay resembles that of the
decay to the $0^+$ states but the energy denominator is up to the third
power.
Being in general this energy of the order of $10~MeV$ this power implies a
factor 100 of suppression for this matrix element\cite{Doi85,Gri92,Suh94}.
The previous equation is rearranged as
\begin{equation}
M_{2\nu}^{GT}(2^+_\sigma) = \sqrt{5}
\sum\limits_{\mu \mu'} (1 \mu 1 \mu'|2 m)
\sum\limits_{N m_1}{}
\mu_N^{-3}  \langle  2^+_\sigma m| \, \Gamma_{\mu'} \, | \ 1^+_N
m_1\rangle \  \langle 1^+_N m_1 \ | \,\Gamma_\mu \, |\	0^+_i \rangle~~.
\end{equation}

Using
\begin{equation}
\mu_N^{-3} = {1\over 2} {\partial^2 \over {\partial E_{J,\sigma }^2}}
\mu_N^{-1}
\end{equation}

\noindent
and the summation method described in\cite{Cas94,Civ93} it is possible to
rewrite the second sum as

\begin{equation}
\begin{array}{l}
\sum\limits_{N m_1}{}{1\over 2} {\partial^2 \over {\partial E_{J,\sigma
}^2}}\{  \mu_N^{-1}
\langle  2^+_\sigma m| \, \Gamma_{\mu'} \, | \ 1^+_N m_1\rangle \
\langle 1^+_N m_1 \
	| \,\Gamma_\mu \, |\  0^+_i \rangle \, \}~=\\
\hspace{2cm}
{1\over 2} {\partial^2 \over {\partial E_{J,\sigma }^2}}
\langle  2^+_\sigma m|
 \sum\limits_{\lambda=1}^{\infty} {{(-1)^{\lambda}} \over
{ E_{J,\sigma }^{\lambda} }}
\Gamma_{\mu'} [ H, [ H, \ldots,[ H,\Gamma_\mu ] \ldots ]^{(\lambda -times)} |
  0^+_i \rangle
\end{array}
\end{equation}

The two body terms of the Hamiltonian (\ref{hamil}) commutes with the
Gamow-Teller operator (\ref{gt}), thus the above multiple commutators are
easy to evaluate. We obtain\cite{Cas94}

\begin{equation}
\begin{array}{ll}
[H,\ldots [H,\Gamma_\mu ]]\ldots ]^{(\lambda -times)}= &
[H_\pi+H_\nu ,\ldots [H_\pi+H_\nu ,\Gamma_\mu]]\ldots ]^{(\lambda -times)}
= \\
& \sum_{\pi \, \nu} \sigma (\pi ,\nu )
[a^{\dagger}_\pi \otimes \tilde a_\nu ]^{1\mu}
\{ \epsilon_\pi -\epsilon_\nu \}^{\lambda}
\end{array}
\end{equation}

\noindent
where $\pi \equiv (\eta_{\pi},l_{\pi},j_{\pi})$ and $\nu \equiv
(\eta_{\nu},l_{\nu},j_{\nu})$. Returning with this expression to the
original formula, Eq. (13), resumming the infinite series and recoupling the
Gamow-Teller operators it is found

\begin{equation}
\begin{array}{ll}
M_{2\nu}^{GT}(2^+_\sigma) =
& \sqrt{5}~{1\over 2} {\partial^2 \over {\partial E_{J,\sigma }^2}}
\left\{ \sum\limits_{\pi\nu,\pi'\nu'}
{{\sigma (\pi ,\nu)\sigma (\pi',\nu') }\over{ E_{J,\sigma }
+\epsilon_\pi -\epsilon_\nu   }}
\langle  2^+_\sigma m| \left[ [a^{\dagger}_\pi \otimes \tilde a_\nu ]^1 \otimes
[a^{\dagger}_{\pi'} \otimes \tilde a_{\nu'}]^1 \right]^{2m}
|\  0^+_i \rangle  \right\} ~=\\
& \sqrt{5}  \sum\limits_{\pi\nu ,\pi' \nu'}
{{\sigma (\pi ,\nu )\sigma (\pi',\nu')}
\over {(E_{J,\sigma }+\epsilon_\pi -\epsilon_\nu)^3} }
 \langle  2^+_\sigma m|
\left[ [a^{\dagger}_\pi \otimes \tilde a_\nu ]^1 \otimes
[a^{\dagger}_{\pi'} \otimes \tilde a_{\nu'}]^1 \right]^{2m}
|\  0^+_i \rangle \label{m11}
\end{array}
\end{equation}

As it was shown in \cite{Cas94} the expression for the nuclear matrix
element of the \bb ~decay to a $0^+$ state is similar to (\ref{m11}), with
a different power in the denominator. Equations (\ref{m11}) and (4.10) of
\cite{Cas94} can be expressed in a compact form as:
\begin{equation}
\begin{array}{ll}
M_{2\nu}^{GT}(J^+_\sigma) =
& \sqrt{J+3}  \sum\limits_{\pi\nu ,\pi' \nu'}
{{\sigma (\pi ,\nu )\sigma (\pi',\nu')}
\over {(E_{J,\sigma }+\epsilon_\pi -\epsilon_\nu)^{J+1}} }
 \langle  J^+_\sigma m|
\left[ [a^{\dagger}_\pi \otimes \tilde a_\nu ]^1 \otimes
[a^{\dagger}_{\pi'} \otimes \tilde a_{\nu'}]^1 \right]^{Jm}
|\  0^+_i \rangle \equiv \\
&  \sum\limits_{\pi\nu ,\pi' \nu'}
{1 \over {(E_{J,\sigma }+\epsilon_\pi -\epsilon_\nu)^{J+1}} }
 \langle  J^+_\sigma m| T^{Jm}(\pi \nu ,\pi' \nu') |\  0^+_i \rangle
\label{m2} \end{array}
\end{equation}

\noindent
For practical purposes the tensor $T^{Jm}(\pi \nu ,\pi' \nu')$ was implicitly
defined in the above equation. The $\sqrt{3}$ for the $J=0$ case comes
from the relation between the scalar product of two vectors and their
coupling to angular momentum zero.

\bigskip
\bigskip
\centerline{\bf The matrix elements $M_{2 \nu}$}
\bigskip
\bigskip
\vskip 0.5pc

We want to evaluate the nuclear matrix element (\ref{m2}) for the \bb
{}~decay of the ground state of \nd , Eq. (\ref{nd}), to the ground and
excited states of \sm ~which are described by the wave functions of Eq.
(\ref{sm}). Each Gamow-Teller operator (\ref{gt}) annihilates a proton and
creates a neutron in the
same oscillator shell and with the same orbital angular momentum.
In the case of the \bb ~of \nd ~it means
that the
operator annihilates two neutrons in the pseudo shell  $\eta_\nu = 5$ and
creates two protons in the abnormal orbit $h_{11/2}$.
As a consequence the only orbitals which in the model space can be connected
through the \bb ~decay are those satisfying
$ \eta_{\pi} = \eta_{\nu} \equiv \eta$, that implies
$l_\pi = l_\nu = \eta$, $j_\nu = \eta - {1 \over 2}$ and
$j_\pi = \eta + {1 \over 2}$.
These are the selection rules described by relations (2) concerning the
change in
occupation numbers. Under this restrictions only one term in the sum
over configurations $\pi \nu ,\pi' \nu'$ survives and thus the
nuclear matrix element $M_{2 \nu}$ (\ref{m2}) can be written as

\begin{equation}
M_{2\nu}^{GT}(J^+_\sigma) = {1 \over {{\cal E}_{J\sigma}^{J+1}} }
\langle J^+_\sigma | T^{Jm}(\pi \nu ,\pi \nu)  | 0^+_i \rangle ,\label{m3}
\end{equation}
where the energy denominator is determined demanding that the Isobaric
Analog State in the intermediate odd-odd nucleus is an eigenstate of the
Hamiltonian (\ref{hamil}). Its
excitation energy is equal to the difference in Coulomb energies $\Delta_C$.
Their expressions are\cite{Cas94}
\begin{equation}
\begin{array}{ll}
	    {\cal E}_{J\sigma} =  &E_{J\sigma} +  \epsilon (\eta_\pi
,l_\pi ,j_\pi = j_\nu + 1) - \epsilon (\eta_\nu ,l_\nu ,j_\nu )=
E_{J\sigma} -\hbar \omega k_\pi 2 j_\pi + \Delta_C .\\
{}~\\
&\Delta_C ={ 0.70 \over A^{1/3}} [2 Z + 1 - 0.76 ( (Z+1)^{4/3} -Z^{4/3} )]
  MeV. \label{den}
\end{array}
\end{equation}

As it was discussed in \cite{Cas94} in the context of the $g.s.
\rightarrow  g.s.$ \bb ~decay Eq. (\ref{m3}) has no free parameters, being
the denominator (\ref{den}) a
well defined quantity. The reduction to only one term comes as a
consequence of the restricted Hilbert proton and
neutron spaces of the model. The initial and final ground states are
strongly correlated with  a very rich structure in
terms of their shell model components.

Since the Hilbert space has been divided in their normal and unique parity
components we need to rearrange the creation an annihilation operators in the
same way, {\em i.e.}

\begin{equation}
\left[ [a^{\dagger}_\pi \otimes \tilde a_\nu ]^1 \otimes
[a^{\dagger}_{\pi} \otimes \tilde a_{\nu}]^1 \right]^{JM} =
\sum\limits_{J_\pi J_\nu} \chi\left\{
\begin{array}{lll}
j_\pi &j_\nu &1\\j_\pi &j_\nu &1\\J_\pi &J_\nu &J
		\end{array}\right\}
\left[ [a^{\dagger}_{\pi} \otimes a^{\dagger}_\pi ]^{J_\pi} \otimes
[\tilde a_\nu \otimes \tilde a_\nu ]^{J_\nu} \right]^{JM}
\end{equation}

The $\chi \{...\}$ is the unitary (Jahn-Hope) 9-j recoupling
coefficient\cite{Var88}. Introducing this expression in (\ref{m3})
together with the explicit form of the wave functions (\ref{nd}) and
(\ref{sm}) we obtain
\begin{equation}
\begin{array}{ll}
M_{2\nu}^{GT}(J^+_\sigma)~ =~
 \sqrt{J+3}  ~\sigma (\pi ,\nu )^2   ~	{\cal E}_{J\sigma}^{-(J+1)}
\sum\limits_{J_\pi J_\nu} \chi\left\{
\begin{array}{lll}
j_\pi &j_\nu &1\\j_\pi &j_\nu &1\\J_\pi &J_\nu &J
		\end{array}\right\}\\
\hspace{2cm}\left[  < (h_{11/2})^6_\pi,  \ J^A_\pi = M^A_\pi = 0|
[a^{\dagger}_{\pi} \otimes a^{\dagger}_\pi ]^{J_\pi}
| \ (h_{11/2})^4_\pi,  \ J^A_\pi = M^A_\pi = 0> \otimes\right. \\
\hspace{2cm} <(12,0)_\pi;  (12,2)_\nu; \ 1 (\lambda ,\mu )_\sigma
K=1 J M |
[\tilde a_\nu \otimes \tilde a_\nu ]^{J_\nu}\\
\hspace{7cm}\left. |(12,0)_\pi; (18,0)_\nu; \ 1 (30,0) K=1 J = M = 0 >
\right]^{JM} \label{m4}
\end{array}
\end{equation}

The matrix element (\ref{m4}) vanishes unless a
pair of protons coupled to total angular momentum zero is created and
two neutrons of the  normal parity space
 coupled to pseudo orbital angular momentum $\tilde L=J$ and
 pseudo spin equal to zero are annihilated.
 The above sum is thus restricted to $J_\pi =0, ~J_\nu = J$.

The operators in the normal space must be recoupled from the $jj-$ to the
$LS-$coupling scheme. The result is
\begin{equation}
[\tilde a_\nu \otimes \tilde a_\nu ]^{J M} =
\sum\limits_{\tilde L \tilde S}
 \chi\left\{
\begin{array}{lll} \tilde l_\nu &1\over 2 &j_\nu\\ \tilde l_\nu &1\over 2
&j_\nu\\ \tilde L &\tilde S &J \end{array}\right\}
[\tilde a_{(0 \tilde\eta )\tilde l ,{1\over 2}}\otimes
\tilde a_{(0\tilde \eta )\tilde l ,{1\over 2}}]^{\tilde L \tilde S}_{JM}
\label{norm}
\end{equation}

The low energy levels are assumed to have pseudo spin $\tilde S = 0$, a
fact that again simplifies the evaluation of the above sum by imposing
$\tilde L = J$.

Using that $j_\pi = j_\nu +1, l_\pi = l_\nu = \eta$ the reduced matrix
elements $\sigma(\pi \nu )$ read
\begin{equation}
\sigma (\pi \nu )^2 = {{8 \eta (\eta + 1)} \over { 3 (2 \eta + 1)}}.
\end{equation}

In the seniority zero approximation the two particle transfer
matrix element of the unique sector of Eq. (22) is evaluated by using the
quasispin formalism and gives\cite{Cas93,Cas94}

\begin{equation}
< j_\pi^{n^A_\pi+2},  \ J^A_\pi = M^A_\pi = 0|
[a^{\dagger}_{\pi} \otimes a^{\dagger}_\pi ]^0
| j_\pi^{n^A_\pi},  \ J^A_\pi = M^A_\pi = 0> =
\left[{{(n^A_\pi+2)(\eta+1-n^A_\pi/2)}\over{\eta +1}}\right]^{1/2}
\end{equation}

The evaluation of the  matrix elements in the normal space of Eq. (\ref{m4})
is performed by using $SU(3)$ Racah calculus to decouple the proton and
neutron normal irreps, and expanding the annihilation operators  of Eq.
(\ref{norm}) in their SU(3) tensorial components. The final result is:
\begin{equation}
\begin{array}{ll}
M_{2\nu}^{GT}(J^+_\sigma)~ =~a(J)~b(n^A_\pi)  ~{\cal E}_{J\sigma}^{-(J+1)}\\
\hspace{1cm}\sum\limits_{(\lambda_0 \mu_0 ) K_0}
<(0 \tilde\eta )1 \tilde l,(0 \tilde\eta )1 \tilde l \| (\lambda_0 \mu_0
) K_0 J>_1 \sum\limits_\rho
<(30,0)1~0,(\lambda_0 \mu_0 )K_0 J\|(\lambda \mu )_\sigma 1 J>_\rho\\
\hspace{1cm}\sum\limits_{\rho'}
\left[\begin{array}{cccc} (12,0) &(0,0) &(12,0) &1\\
(18,0) &(\lambda_0 \mu_0 ) &(12,2) &\rho' \\
(30,0) &(\lambda_0 \mu_0 ) &(\lambda \mu )_\sigma &\rho \\
1 &1 &1 \end{array} \right]
<(12,2)\mid\mid\mid [\tilde a_{0 \tilde \eta),{1\over 2}}
\tilde a_{0 \tilde \eta),{1\over 2}}]^{(\lambda_0 \mu_0 )}
\mid\mid\mid (18,0)>_{\rho'} \label{mgt}
\end{array}
\end{equation}

In the above formula $<..,..\|,,>$ denotes the SU(3) Clebsch-Gordan
coefficients\cite{Dra73}, the symbol  $[...]$ represents a
$9-\lambda\mu$ recoupling coefficient\cite{Mil78}, and
$<..\mid\mid\mid ..\mid\mid\mid ..>$ is the triple reduced matrix
elements\cite{Hir95}.
The energy denominator was defined in (\ref{den}), and

\begin{equation}
\begin{array}{l}
a(0) = {{4\eta}\over{(2\eta + 1)\sqrt{2\eta - 1}}},\hspace{2cm}
a(2) = {2\over {2\eta + 1}}
\left[{ {5 \eta (\eta -1))2\eta -3)}\over{3(2\eta +1)}}\right]^{1/2},\\
b(n^A_\pi ) = [(n^A_\pi + 2)(\eta +1 - n^A_\pi/2)]^{1/2} ~~.
\end{array}
\end{equation}

\bigskip
\bigskip

\centerline{\bf The rotational spectrum of \sm }

\bigskip
\bigskip

The four parameters of the pseudo SU(3) hamiltonian (\ref{hamil}) were
fitted to reproduce the first $2^+$ states in \sm as it was done in
Ref.\cite{Dra84}. Their values are \begin{equation}
\chi = 3.47 eV ~~,~~~\zeta_1= 215 keV ~~,~~~\zeta_2= 50.4 keV ~~.
\end{equation}

The right hand side of Fig. 1 exhibits nine of the lowest energy states
which have been observed in
\sm , grouped in rotational bands. Angular momentum and parity are given for
each level. The left hand side of Fig. 1 shows the calculated spectrum
together with the associated irreps. The gamma band is identified
with K=2.
The general trend is well reproduced but the experimental g.s. band does not
show the rotational structure which is exhibited by the calculated one.
This departure from
an exact rotational behaviour was mentioned in Section 2 and it can be
associated with the relatively small deformation reported for \sm .
In this mass region the deformation
suddenly jumps for $^{152}Sm$ to a value very similar to the deformation of
\nd . The gamma band is present in the g.s.
irrep because both $\lambda$ and $\mu$ are different from zero
and is well fitted.

The excited $0^+$ states are the head of other rotational bands. The
predicted energy gap between them is 125 eV while the experimental
one is 454 eV. These numbers suggest that we have not a clear
identification of
these excited states. This fact has relevance in the study of the
\bb decay to these states.

The BE(2) transition intensities were evaluated using the effective
quadrupole operator\cite{Dra84}

\begin{equation}
Q_0 = e^{eff}_\pi Q_\pi + e^{eff}_\nu Q_\nu~~,~~~~
 e^{eff}_\pi = e + e_{pol}~~,~~~e^{eff}_\nu = e_{pol}
\end{equation}
with $e_{pol} = 0.93 e$. The seniority zero condition imposed to
the nucleons in
abnormal parity orbitals inhibits them to participate in collective
excitations.
This restriction
forces a slightly large value for the polarisation
charge. A similar effect was found in a BE(2) study of
rare earth and actinide nuclei\cite{Dra84}.

The BE(2) intensities of the transition from the first	and second $2^+$ to
the ground state, and from the first $4^+$ to the $2^+$ state are shown in
Table 1 and compared with their experimental values, in Weiskoff units (W.u.).
The agreement is good except for the case of the transition from the second
excited $2^+$ state to the ground state which fails in a factor four.

\bigskip
\bigskip

\centerline{\bf The \bb ~decay of \nd }
\bigskip
\bigskip

In this section we study the two neutrino mode of the double beta decay
\bb ~of
 \nd ~into the ground state, the first excited $2^+$ and the first  and
second excited $0^+$ states of \sm .

In Table 2 the matrix elements, energy denominators,
phase space integrals and predicted half-lives for the \bb ~decay of \nd ~to
the ground state, the first $2^+$ and the first and second excited $0^+$
states of \sm ~are presented.
The matrix elements (26) are given in units of $(m_e c^2)^{-(J+1)}$.
Phase space integrals for the decay to $0^+$ states were evaluated
following
the prescriptions given in\cite{Doi88} with $g_A/g_V=1.0$ and the
kinematical factor for the decay to the $2^+$ state was taken from
\cite{Doi85} renormalized by the above mentioned value of the axial
vector coupling constant. It must be mentioned that these phase space
factors differ in about $10 \%$ with those used in \cite{Cas93,Cas94}
where a different renormalization procedure was used.

As it was mentioned in \cite{Cas94} the predicted half live for the \bb ~
to the ground state of \sm ~is in reasonable agreement with the
experimental data, which vary between $9$ and $17 \times 10^{18}$
years\cite{Moe93,Moe94a,Art93}.

The \bb ~decay to the first excited $0^+$ state is {\em forbidden}.
In this model this is imposed by the fulfilment of an
exact selection rule. It can be understood by
realizing that the pair of annihilation operators $\tilde a_{(0
\tilde 4 ){1\over 2}}$, when expanded in their SU(3) components, have
the couplings $(0,4)\otimes (0,4) = (0,8),(2,4)$ containing a $L=0$ state.
But acting over the \nd ~g.s. irrep (30,0) they cannot couple to the irrep
(20,4) which we associated with the first excited $0^+$ state.	In other words
the transition between members of these particular irreps are forbidden.

The decay to the second excited $0^+$ state is allowed but strongly
cancelled. The reduction of the matrix element by a factor ten, in
comparison with that associated with the decay to the g.s., is partly
related with the fact that though the coupling $(30,0)\otimes (0,8) =
(22,0)$ is
allowed the coupling with $(2,4)$ is forbidden. The predicted
half-life is four orders of magnitude larger than that of the decay to
the g.s.

The \bb ~decay to the $2^+$ state is inhibited by the $\mu_N^3$
dependence of the matrix element as it is discussed in Section 3.
The matrix element of the \bb decay to the first excited $2^+$
state $M_{2\nu}(2^+_{g.s.} ) $ is three orders of magnitude lesser than
the matrix element of the decay to the g.s. $M_{2\nu}(0^+_{g.s.} ) $.

The present results contradicts those previously
published\cite{Pie94,Bar90} were it was speculated that the \bb ~decay of
\nd ~to the first excited $0^+$ state of \sm ~could have a similar intensity
of that to the g.s. We find that in the present formalism  this decay
is forbidden. If we select different
occupation numbers for both \nd ~and \sm , taken the deformation of the
latter nucleus instead of that of the former, we find very similar
results for the decay to the g.s. and the first $2^+$ state, but the
matrix elements of the decay to the first and second $0^+$ states becomes
interchanged with essentially the same values. Considering the difference in
the phase space integrals we predict a half live of the order
$10^{21}$ years for the decay to the first excited $0^+$ state and the
decay to the second excited one becomes forbidden.

The above discussed reduction of the matrix element of the \bb ~decay to
the excited $0^+$ state as compared with the decay to the g.s. is not a
general result of the pseudo SU(3) scheme. A recent analysis of the case of
$^{100}Mo$\cite{Hir95b} shows that both matrix elements are very similar and
that they are
in agreement with the experimental information.
In conclusion, the appearance of selection rules which can produce the
suppression of the matrix elements governing a \bb ~transition is a
consequence of the details of the irreps involved.

The pseudo SU(3) model uses a quite restrictive Hilbert space. The model
could be improved by incorporating mixing between different irreps, via
pairing by example\cite{Tro94}. Also other active shells can be taken
into account in the symplectic extension\cite{Cas92b}. In both cases the
selection rules that impose such strong restrictions to the \bb ~decays
of some nuclei can be superseded.
However if the main part of the wave function is well
represented by the pseudo SU(3) model those forbidden decays will have, in
the better case, matrix elements that will be no greater than $20\%$ of
the allowed ones, resulting in at least one order of magnitude cancellation in
the half-life. In any case these results should be taken into account in the
design of future experiments.

\bigskip
\bigskip

\centerline{\bf Conclusions}

\bigskip
\bigskip

In the present paper we have studied the \bb ~decay mode
of \nd ~to the ground and excited states of \sm .
The transitions have been analyzed in the context of the pseudo SU(3) model.
The experimental spectrum of the g.s. rotational
band of \sm ~was reproduced as
well as the measured half-life of the
\bb ~decay to the g.s.	but the
excited rotational bands were not so well reproduced.
The \bb ~decay to the first excited $0^+$ state was found forbidden in
the model and the decay to the second excited $0^+$ state has a half-life
four orders of magnitude greater that that to the g.s..
The decay to the $2^+$ state is strongly inhibited due to the energy
dependence of the matrix elements $M_{2\nu}(2^+)$,  two powers greater than
that of the matrix element $M_{2\nu}(0^+)$.

It is expected that improving the model
would remove the exact selection rules which forbid some decays.
In any case, if the pseudo SU(3) wave functions are a good representation of
the low-lying energy states of \nd ~ and \sm ,
they will remain inhibited.
\vfill
\eject
\bigskip
\bigskip
\centerline{\bf{Acknowledgement:}}
\bigskip
\bigskip
Work
supported in part by CONACYT under contract 3513-E9310, and by a
CONACYT-CONICET agreement under the project {\em Double Beta Decay.}
O. Civitarese is a fellow of the CONICET, Argentina.

\newpage
\centerline{\bf Table captions}
\vskip 1cm
{\bf Table 1} The BE(2) intensities, in Weiskoff units (W.u.),
of the transition from the first  and second $2^+$ to
the ground state, and from the first $4^+$ to the $2^+$ state of \sm ~
are shown.
The calculated and experimental values are exhibited in the second and
third column, respectively.

\vskip .5cm
{\bf Table 2} The dimensionless \bb ~matrix elements, energy denominators,
phase space integrals and predicted half-lives	for the decay of \nd ~to
the ground state, the first $2^+$ and the first and second excited $0^+$
states of \sm .
\vskip 2cm
\centerline{\bf Figure captions}
\vskip 1cm
{\bf Figure 1} Spectrum of the low-lying states of \sm . The levels are
grouped in rotational bands and they are labeled by angular momentum and
parity. The right hand side contains the experimentally determined levels.
In the left hand side the calculated spectrum is exhibited with the
associated irreps at the bottom.

\newpage
\centerline{Table 1}
\vskip 1cm
$$
\begin{array}{lcc}
transition &theory &experiment \\
\\
\hline \\
2^+_1 \rightarrow 0^+_{g.s.} &55.7 &55.8\\
2^+_2 \rightarrow 0^+_{g.s.} &0.48 &2.0\\
4^+_1 \rightarrow 2^+_1 &78.5 &112.0\\ \\
\hline
\end{array}
$$
\vskip 2cm

\centerline{Table 2}
\vskip 1cm
$$
\begin{array}{llcccc}
\hline
&\vline\\
&\vline &M_{2\nu}^{GT}(J^+_\sigma) &{\cal E}_{J\sigma}[MeV]  &
G_{2\nu}(J^+_\sigma)[yr^{-1}] &\tau^{1/2}_{2\nu}(0^+ \rightarrow
J^+_\sigma)[yr] \\  &\vline \\ \hline  &\vline \\
 0^+\rightarrow 0^+(g.s.) &\vline &.0549 &12.20 &4.94\times 10^{-17}
&6.73\times 10^{18}\\~	&\vline \\
0^+\rightarrow 0^+(1)&\vline &0   &11.83 &5.83\times 10^{-18}
&\infty \\~  &\vline \\
0^+\rightarrow 0^+(2)&\vline &.00499   &11.58 &9.33\times 10^{-19}
&4.31\times 10^{22} \\~  &\vline \\
 0^+\rightarrow 2^+ &\vline &5.38 \times 10^{-5} &12.04 &4.78\times
10^{-17} &7.21 \times 10^{24}\\ ~  &\vline \\ \hline  \\
\end{array}
$$

\end{document}